\documentclass[aps,twocolumn,amsmath,amssymb,preprintnumbers,superscriptaddress,floatfix]{revtex4}

\usepackage{physics}

\usepackage[utf8]{inputenc}
\usepackage{newtxtext}
\usepackage[upint]{newtxmath}
\usepackage{microtype}
\usepackage{textcomp}
\usepackage{dsfont}
\usepackage{bm} 
\usepackage{siunitx}

\usepackage{enumerate}
\usepackage{amsfonts}
\usepackage{amsmath}
\usepackage{amssymb}
\usepackage{color}
\usepackage{soul}

\usepackage{graphicx}

\usepackage[colorlinks,allcolors=blue]{hyperref}
\usepackage{cleveref}

\renewcommand{\propto}{\sim}

\newcommand{\com}[2]{\left[{#1}\,,\,{#2}\right]}                          
\newcommand{\anticomm}[2]{\left\{{#1}\,,\,{#2}\right\}}                    
\newcommand{\covDer}{\bar{\nabla}}                              

\newcommand{\me}[1]{\mathrm{e}^{#1}}                            

\newcommand{\cc}[1]{{#1}^*}                                     
\newcommand{\hc}[1]{{#1}^\dagger}                               
\newcommand{\tc}[1]{\tilde{#1}}                                 

\newcommand{\vect}[1]{\boldsymbol{#1}}                           
\newcommand{\uv}[1]{\vect{e}_{#1}}       

\definecolor{DarkRed}{rgb}{0.80,0,0}
\definecolor{Purple}{rgb}{0.55,0,0.55}

\DeclareMathOperator{\re}{Re}
\DeclareMathOperator{\im}{Im}
\DeclareMathOperator{\sgn}{sgn}
\DeclareMathOperator{\diag}{diag}
\DeclareMathOperator{\antidiag}{antidiag}

\newcommand*{\defeq}{\coloneqq}

\renewcommand{\H}[1]{\hat{#1}}
\newcommand{\V}[1]{\check{#1}}

\newcommand{\up}{\uparrow}                                      
\newcommand{\dn}{\downarrow}                                    

\newcommand{\thouless}{\varepsilon_{\textsc t}}

\let\epsilon\varepsilon

\begin{document}
\title{Controllable Vortex Loops in Superconducting Proximity Systems}
\author{Eirik Holm Fyhn}
\affiliation{Center for Quantum Spintronics, Department of Physics, Norwegian \\ University of Science and Technology, NO-7491 Trondheim, Norway}
\author{Jacob Linder}
\affiliation{Center for Quantum Spintronics, Department of Physics, Norwegian \\ University of Science and Technology, NO-7491 Trondheim, Norway}
\date{\today}
\begin{abstract}
  \noindent 
  Superconducting vortex loops have so far avoided experimental detection despite being the focus of much theoretical work.
  We here propose a method of creating controllable vortex loops in the superconducting condensate arising in a normal metal through the proximity effect.
  We demonstrate both analytically and numerically that superconducting vortex loops emerge when the junction is pierced by a current-carrying insulated wire and give an analytical expression for their radii.
  The vortex loops can readily be tuned big enough to hit the sample surface, making them directly observable through scanning tunneling microscopy.
\end{abstract}
\maketitle
\section{Introduction}
Many key properties of physical systems are determined by topological defects such as dislocations in solids, domain walls in ferroics, vortices in superfluids, magnetic skyrmions in condensed matter systems and cosmic strings in quantum field theories.
In superconductors, the topological entities are vortex lines of quantized magnetic flux.
The topological nature of these vortices makes them stable, which is important for potential applications such as superconducting qubits~\cite{Fedorov2014,Devoret2013,Herr2007}, digital memory and long-range spin transport~\cite{Kim2018}.
Vortices have non-superconducting cores and a phase winding of an integer multiple of $2\pi$ in the superconducting order parameter, leading to circulating supercurrents~\cite{Kwok2016}.

The formation of superconducting vortex loops is topologically allowed, and has theoretically been predicted to form around strong magnetic inclusions inside the superconductor~\cite{Doria2007}, in cylindrically shaped current-carrying superconductors~\cite{samokhvalov1998,samokhvalov1997,kozlov1993} or through vortex cutting and recombination~\cite{Glatz2016,Berdiyorov2018}.
However, no observation of vortex loops in superconducting systems has been found to date.
One challenging aspect is that vortex loops are typically small in conventional superconductors and difficult to stabilize for an extended period of time~\cite{Schonenberger1996}.
Recently it has been shown that vortex loops can be formed in proximity systems by inserting physical barriers, around which the vortices can wrap~\cite{Berdiyorov2018}.

In this manuscript, we present a way to create controllable vortices in mesoscopic proximity systems in a manner which makes them experimentally detectable through scanning tunneling microscopy.
The system considered is a three-dimensional SNS junction pierced by a current-carrying wire which creates the inhomogeneous field responsible for the vortex loops.
In planar SNS-junctions with uniform applied magnetic field, changing the superconducting phase difference between the two superconductors shifts the vortex lines in the vertical direction~\cite{Cuevas2007}.
We here show that the corresponding effect on vortex loops in three dimensions is to change their size.
Thus, these vortex loops are easily tunable.
This makes it possible to make the vortices touch the surface, leaving distinct traces which are directly observable by scanning tunneling spectroscopy~\cite{Stolyarov2018}.

Vortex loops in superconducting systems has previously been predicted using the phenomenological Ginzburg-Landau theory~\cite{Berdiyorov2018,Glatz2016,Doria2007}.
Here we use a fully microscopic framework known as quasiclassical theory and solve the Usadel equation relevant for diffusive systems~\cite{Usadel1970}.
By showing that vortex loop formation occurs in a microscopic theory, we give valuable support to the earlier proposed mechanisms for superconducting vortex loops.
Finally, we discuss how the proposed setup can be realized experimentally.

\begin{figure}
  \centering
  \includegraphics[width=1.0\linewidth]{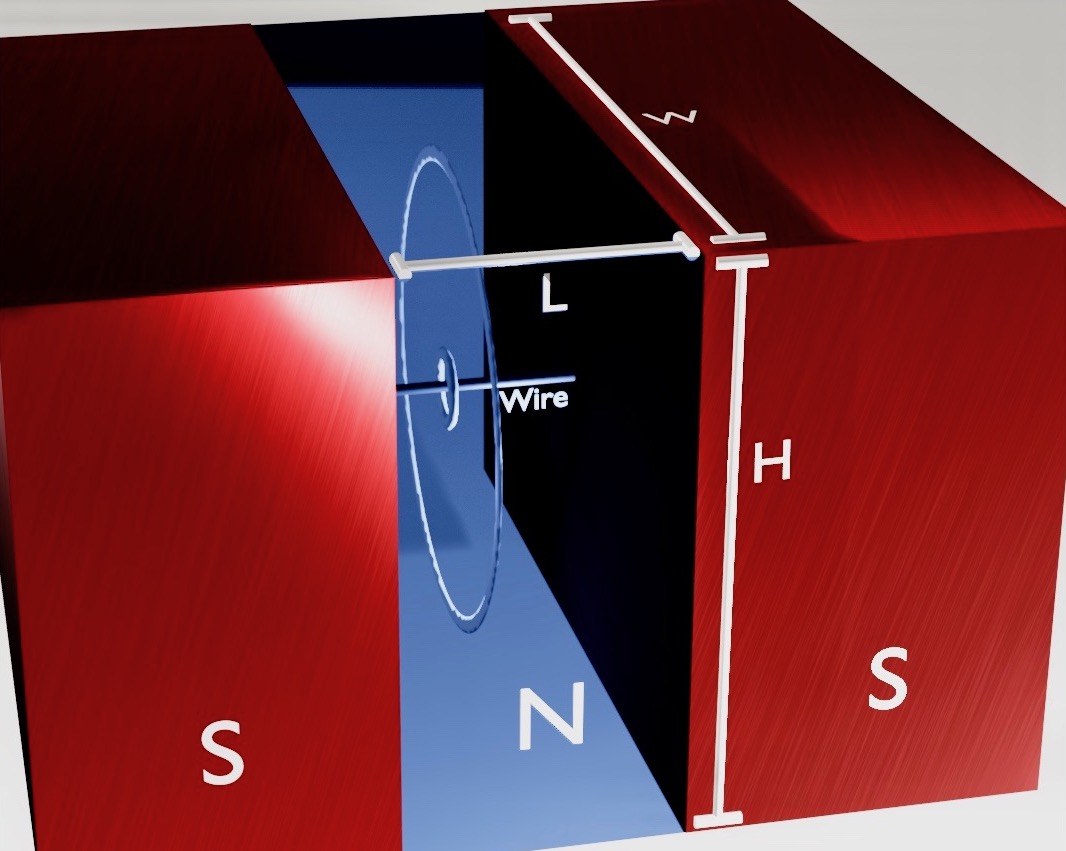}
  \caption{Sketch of three-dimensional SNS junction considered in this manuscript.
    The height, width and length are $H$, $W$ and $L$, respectively, and the junction is pierced by an insulated current-carrying wire.
    Contours of the superconducting vortex loops are shown at the location where they are found in our numerical simulations.
  }
  \label{fig:SNS}
\end{figure}

\section{Methodology}
In this section we discuss the quasiclassical Usadel theory and how it may be used to analyse the SNS-junction depicted in \cref{fig:SNS}.
We first present the mathematical tools and end with the numerical implementation.
\subsection{Quasiclassical Theory}%
\label{sub:quasiclassical_theory}
In the Usadel theory, the system is described by a quasiclassical Green's function from which physical properties can be extracted.
The SNS junction depicted in \cref{fig:SNS} can be treated in the quasiclassical formalism under the assumptions that the Fermi wavelength is much shorter than all other relevant length scales.
If the system is diffusive, meaning that the scattering time is small, the isotropic part dominates and solves the Usadel equation~\cite{Belzig1999,Chandrasekhar2004,Rammer2004,Usadel1970},
which in the normal metal can be written
\begin{align}
  D\covDer\cdot\left(\V{g}\covDer\V{g}\right) + i\com{\epsilon\H{\rho}_3 + \H \Delta}{\V{g}} = 0.
  \label{eq:usadel}
\end{align}
Here, $D$ is a diffusion constant, $\H\rho_3 = \diag(1, 1, -1, -1)$ and $\H{\Delta} = \antidiag(+\Delta,-\Delta,+\Delta^*,-\Delta^*)$ where $\Delta$ is the superconducting gap parameter.
The covariant derivative is $\covDer \V g = \nabla \V g - ie\com{\H \rho_3{\vect A}}{\V g}$, where $e=-\abs{e}$ is the electron charge, $\vect A$ is the vector potential and
\begin{align}
  \V{g} =
  \begin{pmatrix}
    \H{g}^\textsc{r} & \H{g}^\textsc{k} \\
    0   & \H{g}^\textsc{a}
  \end{pmatrix}
  \label{eq:greenFull}
\end{align}
is the quasiclassical impurity-averaged Green's function.
Finally, $(x, y, z) \in [-L/2, L/2]\times [-W/2, W/2]\times [-H/2, H/2]$ in the normal metal.

$\V g$ is normalized such that $\V g \V g = 1$.
We use the convention that when two matrices of different dimensionality is multiplied, the smaller matrix is elevated to the dimensionality of the larger matrix by the tensor product of an identity matrix of the appropriate size.
In equilibrium, the components of the $8 \times 8$ Green's function in \cref{eq:greenFull} are related by the identities $\H{g}^\textsc{k} = \left(\H{g}^\textsc{r} - \H{g}^\textsc{a}\right)\tanh(\varepsilon\beta/2)$ and $\H{g}^\textsc{a} = -\H{\rho}_3 \H{g}^{\textsc{r}\dagger} \H{\rho}_3$, which means that in this case it is sufficient to solve for the retarded component~$\H{g}^\textsc{r}$.

The quasiclassical formalism is not applicable across boundaries because the associated length scale is too short.
The Usadel equation must therefore be solved in the normal metal and superconductors separately, and the solutions must be connected through boundary conditions.
If we assume a low-transparency interface, we may use the Kupriyanov-Lukichev boundary condition
\begin{align}
  \label{eq:bc}
  \zeta_iL_i \uv{n} \cdot (\H g_i^{\textsc{r}} \covDer \H g_i^{\textsc{r}}) = \frac 1 2 \com{\H g_i^{\textsc{r}}}{\H g_j^{\textsc{r}}},
\end{align}
where $\uv n$ is the outward-pointing normal vector for region $i$, $\zeta_i$ is the ratio of the bulk and interface conductances of material $i$ and $L_i$ is the length of material $i$ in the direction of $\uv n$.
For the boundaries interfacing vacuum, $\uv n \cdot \covDer \H g^{\textsc{r}} = 0$.


The Usadel equation can be made dimensionless by introducing the Thouless energy, $\thouless \defeq D/L^2$.
The Usadel equation then becomes dimensionless by doing the substitutions $(x,y,z) \to (x/L, y/L, z/L)$, $\varepsilon \to \varepsilon/\thouless$, $\Delta \to \Delta/\thouless$ and $\bar\nabla \to L\bar\nabla$.

\subsection{Electromagnetic vector potential}%
\label{sub:electromagnetic_gauge}

The magnetic field should satisfy Biot-Savart's law,
\begin{equation}
  \vect B = \frac{\mu}{4\pi}\int \frac{\vect J(\vect r')\times (\vect r - \vect r')}{\abs{\vect r - \vect r'}^3}\dd[3]r',  
  \label{eq:biotSavart}
\end{equation}
where $\mu$ is the permeability and $\vect J$ is the electric current density.
In general, $\vect J$ includes the contribution the induced currents in the normal metal and superconductors in addition to that from the insulated current-carrying wire along the $x$-axis.
However, we will make some assumptions in order to simplify the analytical and numerical calculations.
Firstly, we will assume that the width $W$ and height $H$ is smaller than the Josepshon penetration depth.
In this case we can ignore the screening of the magnetic field by currents inside the normal metal~\cite{Barone1982}.
Secondly, we will neglect the magnetic field produced by the supercurrents produced inside the normal metal.
Thirdly, we will assume that the magnetic field vanish inside the superconductors due to the screening currents.
These last two assumptions is widely used in the context of hybrid structures with constant applied magnetic fields~\cite{Cuevas2007,Alidoust2015,Bergeret2008a}, and has in such conditions been shown to give good agreement with experimental results~\cite{chiodi2012}.
Finally, we will assume that the part of the wire which is inside the superconductors does not contribute to the magnetic field in the normal metal.

The last two assumptions are inaccurate very close to the wire.
Close to the wire the details of screening currents will be important for the magnetic field, but far away we assume that the total contribution from the currents inside the superconductor is zero.
A more precise model could be developed by taking into account screening currents inside the superconductors and solving the Usadel equation self-consistently with Maxwell's equation and the superconducting gap-equation inside the superconductors.
However, we are here interested in the solution far away from the wire, and the details of the magnetic field near the wire should not significantly alter the results.
For this reason we also model the wire as being infinitely thin.

With the assumptions presented above we get a current density which is
\begin{equation}
  \vect J = I \delta(y)\delta(z)\left[\theta(x+L/2) - \theta(x - L/2)\right],
  \label{eq:currDen}
\end{equation}
where $\theta$ is the Heaviside step function.
Inserting \cref{eq:currDen} into \cref{eq:biotSavart} we get
\begin{equation}
  \vect B = \frac{\mu I}{4\pi\rho} \left(\frac{L/2 + x}{\sqrt{(L/2 + x)^2 + \rho^2}} + \frac{L/2 - x}{\sqrt{(L/2 - x)^2 + \rho^2}}\right)\uv \phi
\end{equation}
for $x \in (-L/2, L/2)$, where $\rho = \sqrt{y^2 + z^2}$ and $\uv \phi = (y\uv z - z\uv y)/\rho$.
$\vect B = 0$ for $x < -L/2$ and $x > L/2$.
A vector potential which satisfies $\vect B = \nabla\times \vect A$ is
\begin{widetext}
\begin{equation}
  \vect A = \frac{\mu I }{4\pi} \ln\left( \frac{\sqrt{(L/2 - x)^2 + \rho^2} + L/2 - x}{\sqrt{(L/2 + x)^2 + \rho^2} - L/2 - x} \right)\Bigl[\theta(x+L/2) - \theta(x - L/2)\Bigr]\uv x,
  \label{eq:vectPot}
\end{equation}
\end{widetext}
as can be seen from insertion or calculated directly from Biot-Savarts law by using that $\nabla\times (\vect J(\vect r')/\abs{\vect r - \vect r'}) = \vect J(\vect r')\times (\vect r' - \vect r) /\abs{\vect r' - \vect r}^3$.

\subsection{The Ricatti Parametrization}
In the Ricatti parametrization~\cite{schopohl_arxiv_98} of $\H g^{\textsc r}$, the parameter is the $2\times 2$ matrix $\gamma$ and the retarded Green's function is written
\begin{align}
  \H g^{\textsc r} =
  \begin{pmatrix}
    N & 0 \\
    0 & -\tc N
  \end{pmatrix}
  \begin{pmatrix}
    1 + \gamma \tc \gamma & 2\gamma \\
    2\tc\gamma & 1+ \tc\gamma\gamma
  \end{pmatrix},
\end{align}
where $N \defeq \left(1 - \gamma\tc\gamma\right)^{-1}$ and tilde conjugation is $\tc\gamma(\varepsilon) = \cc\gamma(-\varepsilon)$.

Since the superconducting correlations in our system are spin-singlet, we may write $\gamma_{\textsc{n}} = \antidiag(a, -a)$ and $\gamma_{\textsc{bcs}} = \antidiag(b, -b)$,
where $\gamma_{\textsc{n}}$ and $\gamma_{\textsc{bcs}}$ are the Ricatti parameters in the normal metal and superconductors, respectively.
Substituting this into \cref{eq:usadel,eq:bc} we obtain the dimensionless equations
\begin{align}
\label{eq:usadelRicatti}
\begin{split}
  \nabla^2 a = \frac{2\tc a \nabla a \cdot \nabla a}{1 + a\tc a}  + \frac{4(1 - a \tc a)Le\vect A\cdot (aLe\vect A + i\nabla a)}{1 + a\tc a}\\
  + 2iLe(\nabla\cdot \vect A)a- 2i\varepsilon a,
\end{split}
\end{align}
and
\begin{align}
  \begin{split}
  \uv n \cdot \nabla a = 
  \frac{(1 + a\tc b)(b - a)}{\zeta (b\tc b + 1)}
  + 2ia \uv n \cdot \vect A eL,
  \end{split}
  \label{eq:bcRicatti}
\end{align}
where $L$ is the length which is used to define the Thouless energy, $\thouless$.
The corresponding equations for $\tc a$ and $\uv n \cdot \nabla \tc a$ is found by tilde conjugating \cref{eq:usadelRicatti,eq:bcRicatti}.

\subsection{Observables}
As mentioned initially, a vortex is accompanied by a non-superconducting core and a circulating supercurrent. 
Both the superconducting order parameter and the supercurrent can be extracted from the quasiclassical Green's function.
In the following it will be useful to write
\begin{equation}
  \H g^{\textsc r} =
  \begin{pmatrix}
    g & f \\
    -\tc f & -\tc g
  \end{pmatrix}.
  \label{eq:gr}
\end{equation}
There are only singlet correlations in the SNS system, so $f = \antidiag(f_s, -f_s)$.

The local density of states for spin-band $\sigma$ at energy $\varepsilon$ and location $\vect r$ can be written
\begin{equation}
  N_\sigma(\varepsilon, \vect r) = N_0\real\{g_{\sigma\sigma}(\varepsilon, \vect r)\},
  \label{eq:DOS}
\end{equation}
where $N_0$ is the normal state density of state at the Fermi surface.
In the normal metal we can write \cref{eq:DOS} in terms of $a$,
\begin{equation}
  N(\varepsilon, \vect r) \defeq \frac{N_\up(\varepsilon, \vect r) + N_\dn(\varepsilon, \vect r)}{2} = N_0 \frac{1 - a \tc a}{1 + a \tc a}.
  \label{eq:DOSHM}
\end{equation}
In the cores of vortices we expect $N = N_0$ for all energies, which happens when $a(\varepsilon) \equiv 0$.


The current density is~\cite{Belzig1999}
\begin{equation}
  \vect j = \frac{N_0 eD}{4} \int_{-\infty}^{\infty}\Trace\left(\H\rho_3 \left[\V g \covDer\V g\right]^{\textsc k}\right) \dd{\varepsilon}.
  \label{eq:currentGen}
\end{equation}
Inserting \cref{eq:gr}, using the relations $\H{g}^\textsc{a} = -\H{\rho}_3 \H{g}^{\textsc{r}\dagger} \H{\rho}_3$, $\H{g}^\textsc{k} = \left(\H{g}^\textsc{r} - \H{g}^\textsc{a}\right)\tanh(\varepsilon\beta/2)$,
\cref{eq:currentGen} can be rewritten
\begin{align}
  \begin{split}
  \vect j = \frac{N_0 eD}{2} \int_{-\infty}^{\infty}\tanh\left(\frac{\beta\varepsilon}{2}\right)\Trace\Bigl(\re\left[\hc{\tc f}\nabla \hc f - f \nabla\tc f\right] \\
  + 2e\vect A \im\left[f\tc f - \hc{\tc f}\hc f\right]\Bigr) \dd{\varepsilon}.
  \end{split}
\end{align}


Written in terms of the quasiclassical Green's function, the superconducting order parameter is
\begin{multline}
  \Psi(\vect r) \defeq \expval{\psi_\uparrow(\vect r, 0)\psi_{\downarrow}(\vect r, 0)} \\
  = \frac{N_0}{2}\int_{-\infty}^{\infty}f_{s}(\vect r, \varepsilon)\tanh(\varepsilon\beta/2)\dd{\varepsilon}.
\end{multline}
where $\psi_{\sigma}(\vect r, t)$ is the field operator which destroys an electron with spin $\sigma$ at position $\vect r$ and time $t$, $N_0$ is the normal state density of states and $\beta = 1/k_{\textsc b}T$.

\subsection{Numerics}
The Usadel equation was solved numerically using a finite element scheme.
See for instance~\cite{Amundsen2016} to see how to set up solve the nonlinear Usadel equations in a finite element scheme by the use of the Newton-Rhapson method.
The program was written in Julia~\cite{Bezanson2017}, we used linear hexehedral elements and JuAFEM.jl~\cite{Carlsson2019} was used to iterate through the cells.
Gauss-Legandre quadrature rules of fourth order was used to integrate through the cells and Romberg integration was used to integrate over energy.
See for instance~\cite{sauer2013}.
Finally, forward-mode automatic differentiation~\cite{RevelsLubinPapamarkou2016} was used to calculate the Jacobian.

\section{Results and Discussion}
Here we present first an analytical solution of the Usadel equation in the weak proximity effect regime, then we show numerically that the findings are also present also in the full proximity effect regime.
Dimensionless quantities are used in the analytics with distances being measured relative to the length of the half-metal, $L$, and energies being measures relative to the Thouless energy $\thouless = D/L^2$, where $D$ is the diffusion constant in the half-metal.
\subsection{Analytics}%
\label{sub:analytics}
Before solving the Usadel equation we must determine the solution in the superconducors.
We will show that it suffices to use the bulk solution
\begin{equation}
  \H g_{\textsc{bcs}} = \left[\frac{\theta\left(\varepsilon^2 - \abs{\Delta}^2\right)}{\sqrt{\varepsilon^2 - \abs{\Delta}^2}}\sgn(\varepsilon) - \frac{\theta\left(\abs{\Delta}^2 - \varepsilon^2\right)}{\sqrt{\abs{\Delta}^2 - \varepsilon^2}}i \right]\left(\varepsilon\H\rho_3 + \H\Delta\right),
  \label{eq:bulkBCS-hm}
\end{equation}
in the superconductors when a certain condition is fulfilled. Let $\lambda$ (to be defined quantitatively below) be the length-scale over which the Green function recovers its bulk value in the superconductor. The criterion for neglecting the inverse proximity effect in the superconductors is then that the normal-state conductance of the superconductors for a sample of length $\lambda$ is much larger than the interface conductance and that the length of each superconductor is not small compared to $\lambda$. We now proceed to prove this.

The vector potential,~\eqref{eq:vectPot}, is zero inside the superconductors, so the Usadel equation simplifies to
\begin{align}
  D_{\textsc{sc}}\nabla\cdot\left(\H{g}^{\textsc{r}}\nabla\H{g}^{\textsc{r}}\right) + i\com{\epsilon\H{\rho}_3 + \H{\Delta}}{\H{g}^{\textsc{r}}} = 0
  \label{eq:usadel-hm}
\end{align}
in the superconductor at $x<-1/2$.
To show that we can use the bulk solution in the limit $L_{\textsc{sc}} \to \infty$, let
\begin{equation}
  \H g^{\textsc{r}} = \H g_{\textsc{bcs}} + \delta \H g
\end{equation}
This gives an equation for $\delta \H g$,
\begin{align}
  D_{\textsc{sc}}\nabla\cdot\left(\left[\H g_{\textsc{bcs}} + \delta \H g\right]\nabla\delta\H{g}\right) + i\com{\epsilon\H{\rho}_3 + \H{\Delta}}{\delta\H{g}} = 0,
  \label{eq:dg}
\end{align}
where we have used that $\H g_{\textsc{bcs}}$ solves the \cref{eq:usadel-hm} for a bulk superconductor.
Next, assume the inverse proximity effect to be weak, such that $\delta \H g \ll \H g_{\textsc{bcs}}$.
Using that $\H g_{\textsc{bcs}}\H g_{\textsc{bcs}}=1$, this yields
\begin{align}
  D_{\textsc{sc}}\nabla^2\delta\H{g} + i\H g_{\textsc{bcs}}\com{\epsilon\H{\rho}_3 + \H{\Delta}}{\delta\H{g}} = 0.
  \label{eq:dg2}
\end{align}
$\H g_{\textsc{bcs}} + \delta \H g$ must also satisfy the normalization condition $(\H g^{\textsc{r}})^2 = 1$, so
\begin{equation}
  \left(\H g_{\textsc{bcs}} + \delta \H g\right)^2 = 1 \implies \anticomm{\H g_{\textsc{bcs}}}{\delta\H g} = 0.
\end{equation}
Hence, using that $\com{\epsilon\H{\rho}_3 + \H{\Delta}}{\H g_{\textsc{bcs}}} = 0$,
\begin{align}
  \H g_{\textsc{bcs}}\com{\epsilon\H{\rho}_3 + \H{\Delta}}{\delta\H{g}} &=
  (\epsilon\H{\rho}_3 + \H{\Delta})\H g_{\textsc{bcs}}\delta\H{g} 
  + \delta\H{g}(\epsilon\H{\rho}_3 + \H{\Delta})\H g_{\textsc{bcs}} \notag \\
  &= \anticomm{\delta\H g}{(\epsilon\H{\rho}_3 + \H{\Delta})\H g_{\textsc{bcs}}}.
\end{align}
Finally, from
\begin{equation}
  \left(\epsilon\H{\rho}_3 + \H{\Delta}\right)^2 = \varepsilon^2 - \Delta^2
\end{equation}
we get that $\delta\H g$ is an eigenfunction of the Laplacian,
\begin{equation}
  \nabla^2 \delta\H g = \lambda^{-2} \delta \H g
\end{equation}
where
\begin{align}
  \lambda^{-2} = -\frac{2i}{D_{\textsc{sc}}}\Biggl[\sgn(\varepsilon)\sqrt{\varepsilon^2 - \abs{\Delta}^2}\theta\left(\varepsilon^2 - \abs{\Delta}^2\right) \notag
  \\+ i\sqrt{\abs{\Delta}^2 - \varepsilon^2}\theta\left(\abs{\Delta}^2 - \varepsilon^2\right)\Biggr].
  \label{eq:lambda}
\end{align}
We can choose the sign of $\lambda$ to be such that $\real(\lambda) > 0$.

Let $L_{\textsc{sc}}$ be the length of the superconductor in multiples of the length of the normal metal.
Using the boundary condition
\begin{equation}
  \nabla \delta\H g\bigr\rvert_{\vect r \in \Omega} = 0,
\end{equation}
where $\Omega$ is the boundary not interfacing the normal metal, we get
\begin{equation}
  \delta \H g(\varepsilon, x, y) = C\left[ \me{-\abs{x + 1/2}/\lambda} + \me{-2L_{\textsc{sc}}/\lambda + \abs{x + 1/2}/\lambda}\right],
  \label{eq:deltag_sol}
\end{equation}
where $C$ is some a function of $y$ and $\varepsilon$ to be determined by the final boundary condition.
From the remaining boundary condition, \cref{eq:bc}, we get
\begin{equation}
  C = \frac{\lambda\H g_{\textsc{bcs}} \com{\H g_{\textsc{bcs}} + \delta \H g}{\H g_{\textsc{n}}}}{2\left(1 - \me{-2 L_{\textsc{sc}}/\lambda}\right)\zeta_{\textsc{sc}}L_{\textsc{sc}}}.
  \label{eq:matCurr}
\end{equation}
From \cref{eq:deltag_sol} we see that $\real (\lambda)$ can be interpreted as the penetration depth of $\delta g$.
Note that $\real(\lambda)$ is bounded by including the effect of inelastic scattering, which is done by the substitution $\varepsilon \to \varepsilon + i\delta$ for some positive scattering rate $\delta$~\cite{Dynes1984}.
This ensures that $1/\left(1 - \me{-2L_{\textsc{sc}}/\lambda}\right)$ remains finite as $\epsilon \to \Delta$.
Thus, we see from \cref{eq:matCurr} that $C$, and therefore $\delta g$, becomes negligble when
\begin{equation}
  \zeta_{\textsc{sc}}L_{\textsc{sc}}/\real(\lambda) \gg 1,
  \label{eq:inverseCriterion}
\end{equation}
provided that the length of the superconductor $L_\textsc{sc}$ is not small compared to the maximal penetration depth, $\max[\real(\lambda)]$.

$\zeta_{\textsc{sc}}$ is proportional to the conductance of the whole superconductor and therefore also with $1/L_\textsc{sc}$.
Therefore, $\zeta_{\textsc{sc}}L_{\textsc{sc}}/\real(\lambda)$ is the ratio of the normal-state conductance of a superconductor of length $\real(\lambda)$ to the interface conductance. 
Taking the superconducting coherence length $\xi$ as a measure of the inverse proximity effect penetration depth $\real(\lambda)$, we see that the criterion~\cref{eq:inverseCriterion} is indeed experimentally feasible. The equation is fulfilled for a low-transparency interface and for a superconductor that is larger than the coherence length.
A similar calculation shows that we can use $\H g_{\textsc{bcs}}$ also in the superconductor at $x > 1/2$.

Solving for the Ricatti parameter in the superconductors we get that $\gamma_{\textsc{bcs}} = \antidiag(b, -b)$ with
\begin{widetext}
\begin{equation}
  b =
    \frac{\Delta}{\varepsilon + i\sqrt{\abs{\Delta}^2 - \varepsilon^2}}\theta(\abs{\Delta} - \abs{\varepsilon}) +
    \frac{\Delta\sgn(\varepsilon)}{\abs{\varepsilon} + \sqrt{\varepsilon^2 - \abs{\Delta}^2}}\theta\left(\abs{\varepsilon} - \abs{\Delta}\right). 
\end{equation}
\end{widetext}

The non-linear Usadel equation does not have a general analytical solution, but it can be solved analytically in an approximate manner far away from the wire.
If we assume the proximity effect to be weak, we can keep only terms which are linear in $a$, $\tc a$ and their gradients.
In this case the Usadel equation~\eqref{eq:usadelRicatti} decouples:
\begin{align}
  \label{eq:usadelLinearized}
  \nabla^2 a = 4eL\vect A\cdot (aeL\vect A + i\nabla a) + 2ieL(\nabla\cdot \vect A) a - 2i\varepsilon a.
\end{align}
\Cref{eq:usadelLinearized} can be further simplified when we only consider regions where $\rho\gg 1$, with $\rho = \sqrt{y^2 + z^2}$.
The solution of \cref{eq:usadelLinearized} is constant in $y$ and $z$ when $\vect A = \vect 0$, and by assuming this is also approximately true when $\abs{eL\vect A} \ll 1$, we can neglect the terms $\partial_y^2 a$ and $\partial_z^2 a$.
Finally, we can  simplify the calculations further by Taylor expanding the vector potential,
\begin{equation}
  L e\vect A = -n\pi \frac 1 \rho \uv x + \mathcal{O}\left(\frac{1}{\rho^2}\right)\uv x,
\end{equation}
where
\begin{equation}
  n = -\frac{eL\mu I}{4\pi^2}.
\end{equation}
We keep only the first term in the Taylor expansion. 

\Cref{eq:usadelLinearized} can now be solved exactly, and by applying the linearized boundary conditions,
\begin{align}
\label{eq:nBcLinearized}
  \uv n \cdot \nabla a &= 
  \frac{(b + a[b\tc b - 1])}{\zeta(b\tc b + 1)}
  + 2ia \uv n \cdot \vect A eL,
\end{align}
the solution can be written on the form
\begin{widetext}
\begin{align}
    a = \frac{c\me{i\phi_l + u(x-0.5)}}{(k - d)^2\me{k} - (k + d)^2\me{-k}}
    \Bigl\{(k - d)\left(\me{k(x-0.5)} + \me{i\delta\phi - u} \me{-k(x+0.5)}\right)
    +(k + d)\left(\me{k(0.5-x)} + \me{i\delta\phi - u}\me{k (x+0.5)}\right) \Bigr\},
  \label{eq:linSol}
\end{align}
\end{widetext}
where
\begin{align}
  &\delta\phi = \phi_r - \phi_l, \\
  &c = \frac{\abs{b}}{\zeta (b\tc b + 1)},
  \qquad \text{   } \qquad
  d = \frac{(b\tc b - 1)}{\zeta (b\tc b + 1)},
  \\
  &u = - \frac{2\pi i n}{\rho}
  \quad
  \qquad \text{and} \qquad
  k = \sqrt{-2i\varepsilon}.
\end{align}
From \cref{eq:linSol} we see that $a$ vanishes at $x = 0$ and $i\delta\phi - u = i(2N+1)\pi$, where $N$ is any integer.
This happens at
\begin{equation}
  \rho = \frac{2n}{1 + 2N - \frac{\phi_r - \phi_l}{\pi}}
  \label{eq:vortPosR}
\end{equation}
This means that $f$ and hence also $\Psi$ vanish at these points.
By Taylor expanding $a$ to first order around a root located at $(0, \tilde \rho)$ we find
\begin{equation}
  a \propto B_1\cos(\theta + \alpha_1) + iB_2\cos(\theta+\alpha_2),
\end{equation}
where $x \propto \cos\theta$ and $ \rho-\tilde \rho \propto \sin\theta$, $B_1^2 = 5\abs{k}^2/4 - \abs{k}d + 2d^2$, $B_2^2 = \abs{k}^2/4 + d^2$, $\alpha_1 = \tan^{-1}[(\abs{k}/2 + d)/(\abs{k} - d)]$ and $\alpha_2 = \tan^{-1}(\abs{k}/2d)$.
Hence, these roots have a phase winding of $2\pi$, as is characteristic for vortices.
\Cref{eq:vortPosR} is our main analytical result as it predicts how the radius of the vortex loops depends on the tunable parameters of the system: the current through the wire and the applied phase difference. Although it was obtained using approximations, we demonstrate below that it matches the full numerical solution of the exact Usadel equation very well.

Note that the radius, $\rho$, of the largest vortex loop given \cref{eq:vortPosR} can be made arbitrary large by letting $\phi_r -\phi_l$ approach $\pi$.
Thus, for a given sample size $L\times W\times H$ and current $I$, there is a superconducting phase difference for which the vortex loop hits the surface and can be directly detected experimentally.

It is expected that a change in the superconducting phase difference will change the radii of vortex loops.
This is because changing the phase difference is equivalent to changing the applied supercurrent through the junction.
The applied current will be deflected by the circulating currents associated with the vortices and hence produce a reactionary force on the vortices.
See for instance~\cite{sonin1997}.
What is more surprising, however, is that changing the superconducting phase difference can make the vortices arbitrarily large so that they can always be made to hit the surface.
If this feature is gerenally true for other systems with vortex loops it could prove useful for the study of systems containing vortex loops which are less obviously controllable than the one considered in the present manuscript, but which are easier to design in a lab.
For instance, one possibility is to grow the normal metal around a magnetic dipole.
Ref.~\cite{Doria2007} found that vortex loops can form around magnetic dipole inclusions in superconductors if the magnetic field is strong enough, so thre are reasons to believe that vortex loops can also form around magnetic dipoles embedded in a SNS-junction.
The magnetic field from a dipole can, unlike the magnetic field from a wire, not be altered in strength.
Nevertheless, if the field is strong enough to produce vortices, altering the superconducting phase difference could be a way to increase the size of the vortex to the point where it touches the surface and becomes directly observable.

\subsection{Numerics}%
\label{sub:numerics}
We now proceed to show numerical results in the full (non-linear) proximity effect regime.
We have set the parameters $\abs{\Delta} = 4\thouless$, $\zeta = 3$, $W = H = 6L$ and $\phi_l = 0$ common for all the numerical calculations.
We include the effect of inelastic scattering by doing the substitution $\varepsilon \to \varepsilon + i\delta$ where $\delta = 0.001\abs{\Delta}$ in order to avoid the divergence of $\H g_{\textsc{bcs}}$ at $\varepsilon = \abs{\Delta}$~\cite{Dynes1984}.
\begin{figure}
  \centering
  \includegraphics[width=1.0\columnwidth]{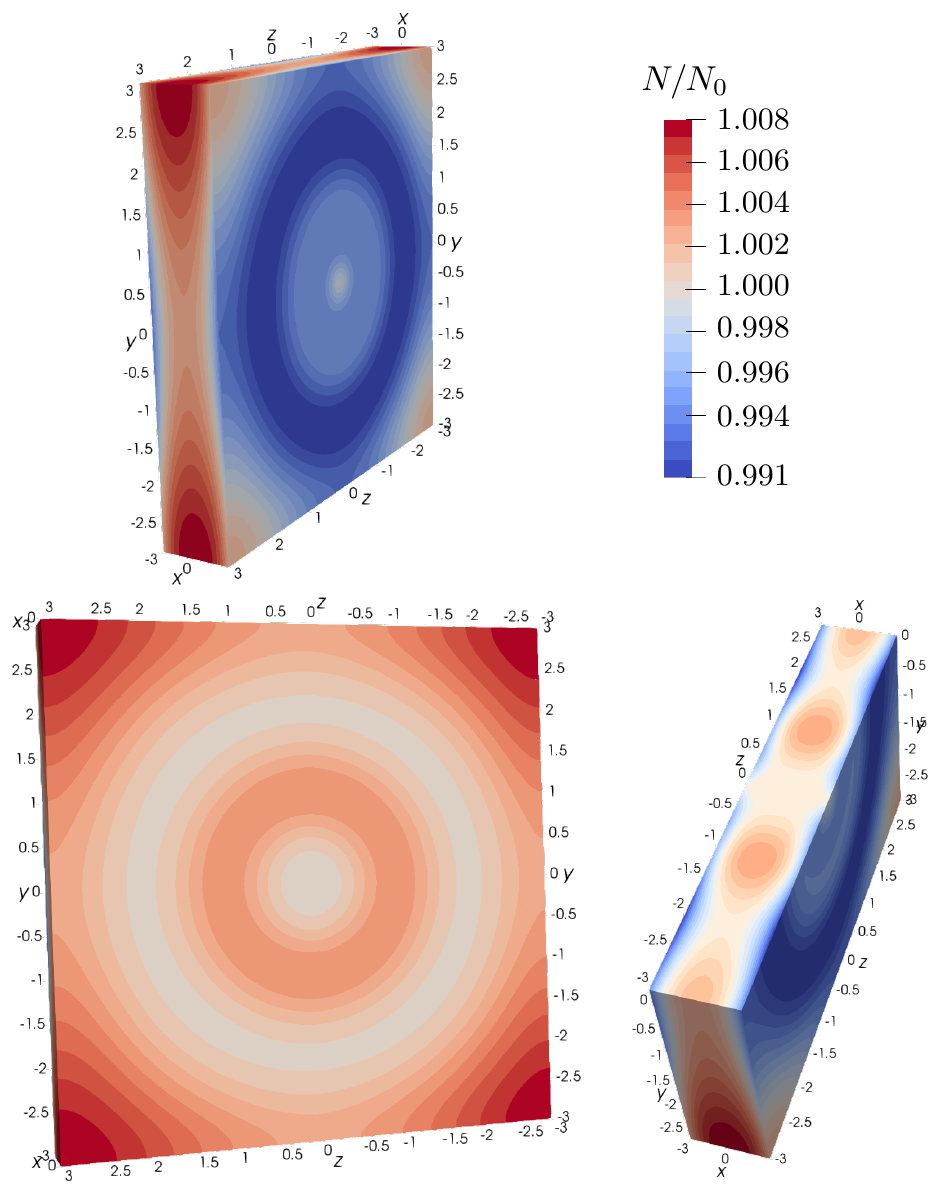}
  \caption{Local density of states $N$ relative to the normal state density of states $N_0$ at energy $\varepsilon = 0.5\abs{\Delta}$, where $\Delta$ is the superconducting gap parameter.
      The lower left shows a cross section at $x = 0$ and the lower right shows a cross section at $y = 0$.
  Here $n=1$ and the superconducting phase difference is $\phi_r=0$.}
    \label{fig:DOS}
\end{figure}
\begin{figure}
  \centering
  \includegraphics[width=1.0\columnwidth]{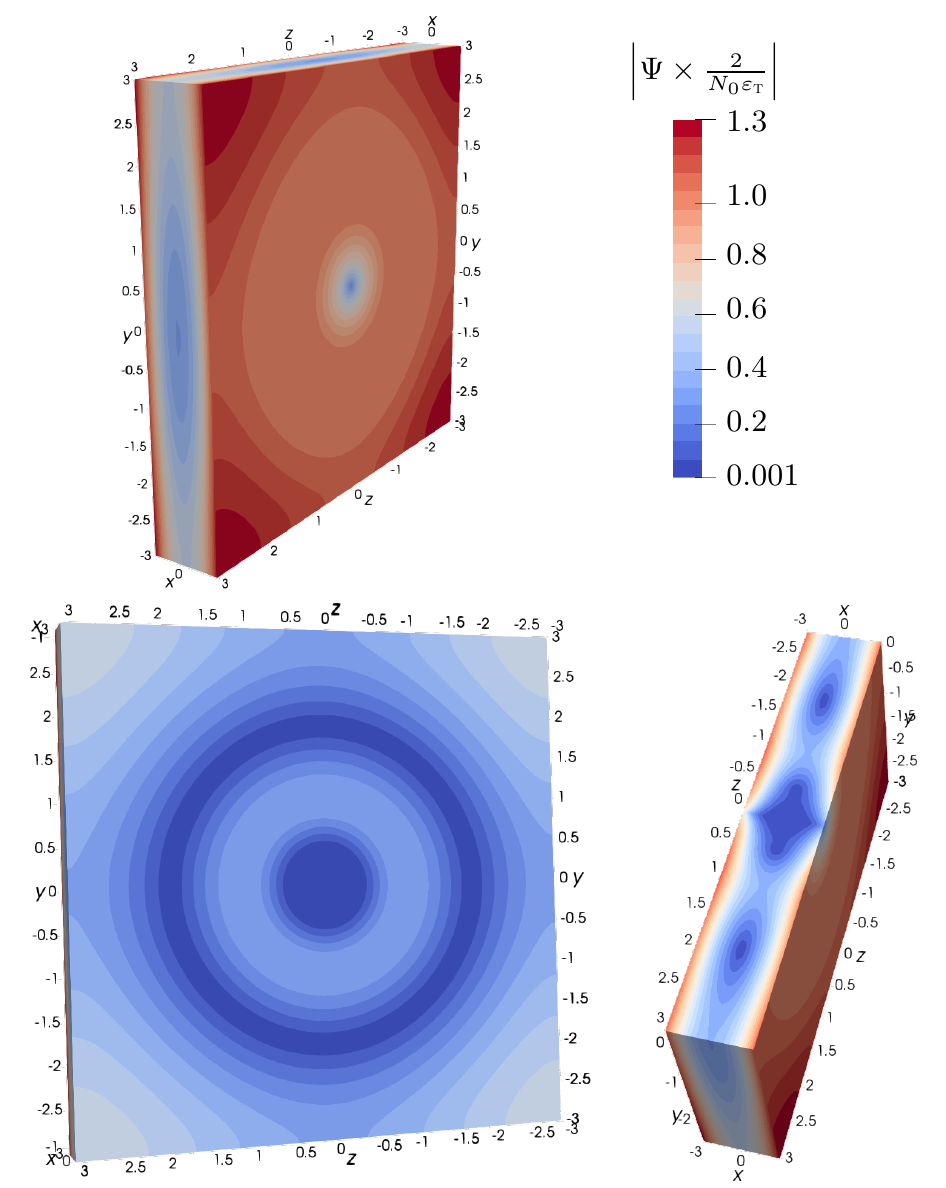}
    \caption{Amplitude of the superconducting order parameter $\Psi$ for $n=1$ and superconducting phase difference  $\phi_r = 0$.
    The lower left shows a cross section at $x = 0$ and the lower right shows a cross section at $y = 0$.}
    \label{fig:psi}
\end{figure}
\begin{figure}
  \centering
  \includegraphics[width=1.0\columnwidth]{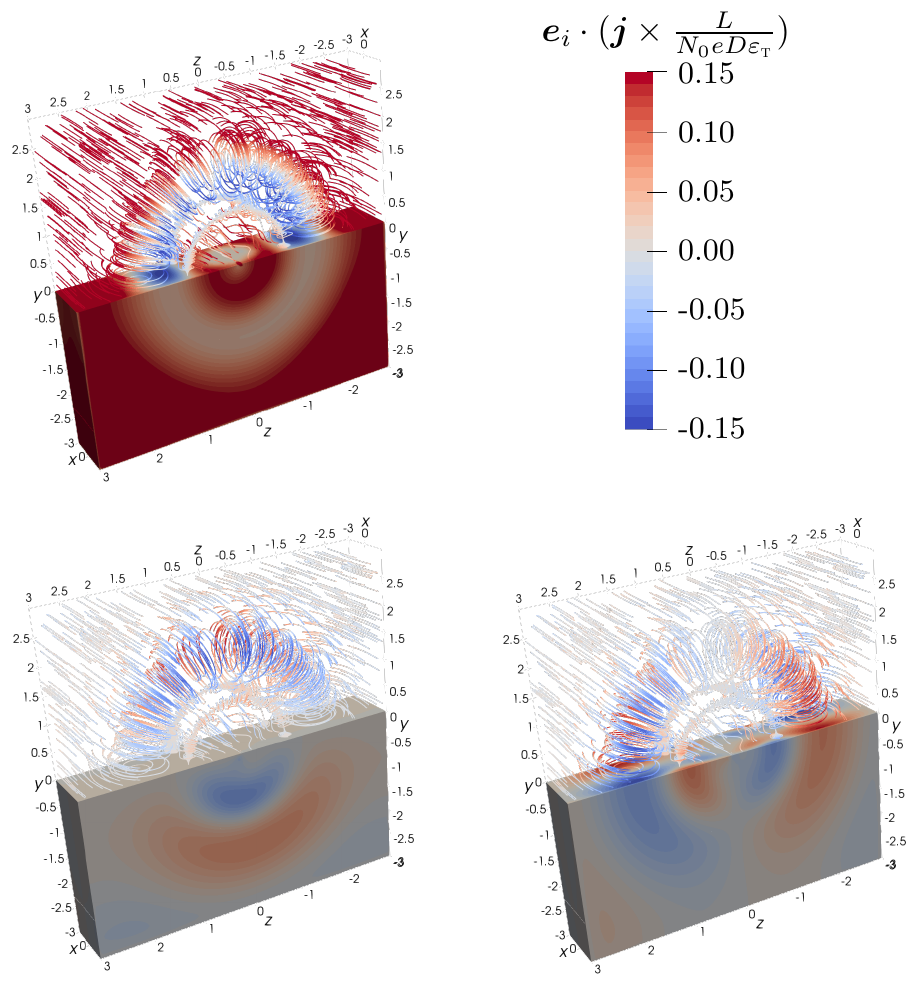}
  \caption{Plot of the three different components of the supercurrent, $\uv x \cdot \vect j$ (upper left), $\uv y \cdot \vect j$ (lower left) and $\uv z\cdot \vect j$ (lower right).
    The lower half shows the value of the current on the surface in color and the upper half shows streamlines of the current with the current strength indicated by the same color.
    Here $n=1$ and $\phi_r=0$.}
    \label{fig:curr}
\end{figure}
\begin{figure}
  \centering
  \includegraphics[width=1.0\linewidth]{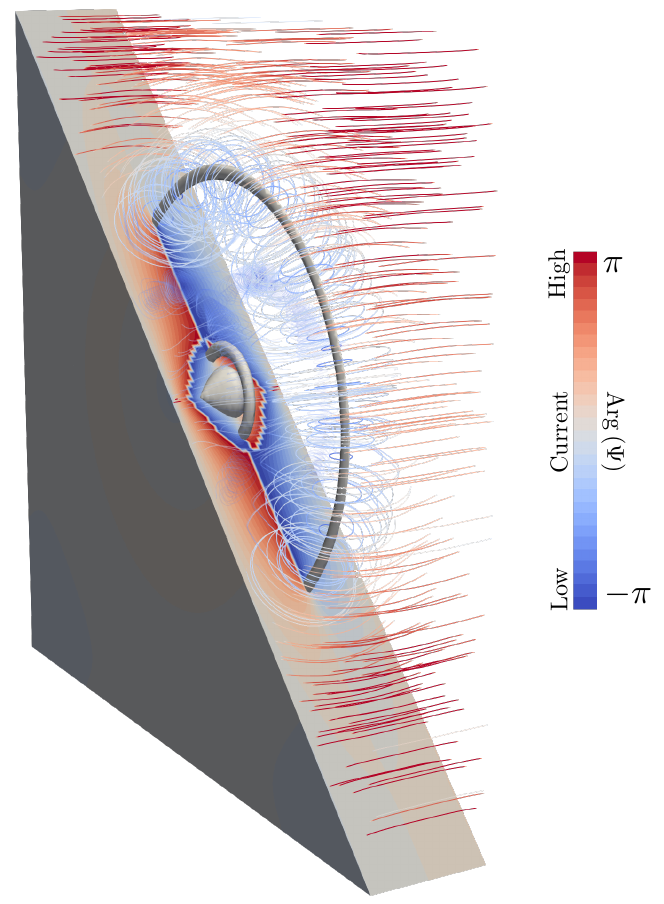}
    \caption{Plot of the phase of the superconducting order parameter $\Psi$ on a the surface of a diagonally cut part of the normal metal, contour plot of its amplitude, $\abs{\Psi}$, and streamlines of the supercurrent $\vect j$. Here $n = 1$ and $\phi_r = 0$.}
    \label{fig:combined}
\end{figure}

\begin{figure*}
  \centering
  \includegraphics[width=1.0\linewidth]{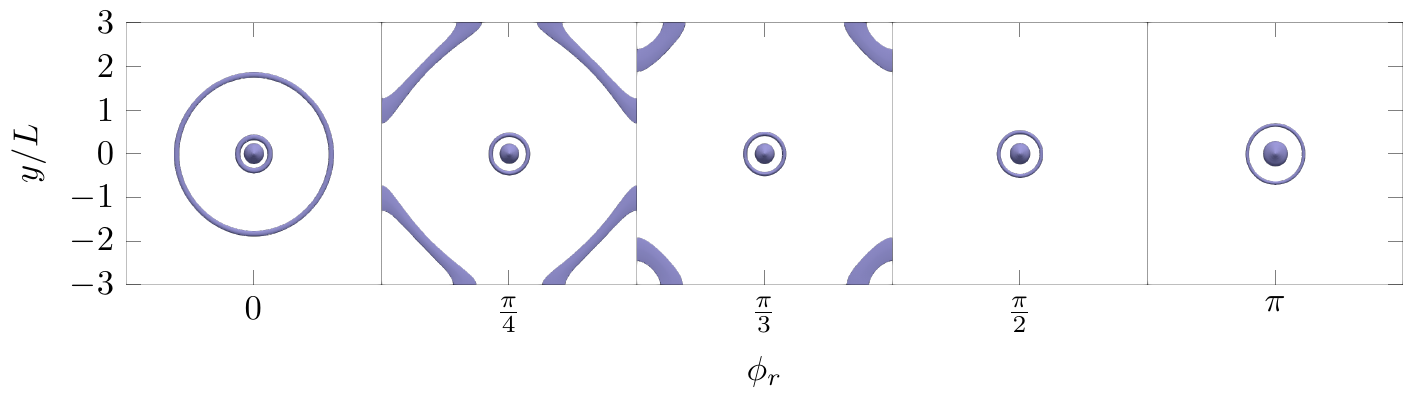}
    \caption{Contour plot of the amplitude of the superconducting order parameter $\Psi$ for $n=1$ and various values of the superconducting phase difference $\phi_r$.}
  \label{fig:loop_phi}
\end{figure*}
\begin{figure*}
  \centering
  \includegraphics[width=1.0\linewidth]{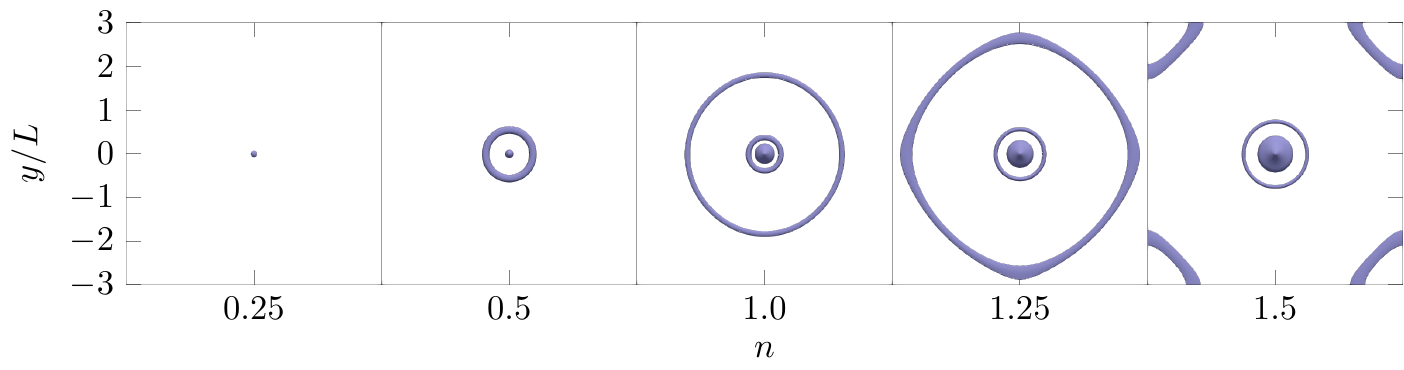}
  \caption{Contour plot of the amplitude of the superconducting order parameter $\Psi$ for superconducting phase difference $\phi_r=0$ and various values of $n$.}
  \label{fig:loop_I}
\end{figure*}

Numerically we find that vortex loops do indeed form at the locations predicted by the analysis.
There are circular paths around the origin where the superconducting order parameter vanish and the local density of states is equal to that of the normal state.
This can be seen in \cref{fig:DOS,fig:psi} which shows the local density of states and the amplitude of the Cooper pair correlation function $\Psi$, respectively.
Around these loops there are a circulating supercurrent, as can be seen in \cref{fig:curr}, and a phase winding in the order parameter of $2\pi$.
\Cref{fig:combined} shows a contour plot of $\abs{\Psi}$, which shows the location of the vortices, together with the circulating supercurrent $\vect j$ as well as the phase of $\Psi$, which shows that there is indeed a phase winding of $2\pi$ around the vortices.

We find that the positions of the vortex loops match with \cref{eq:vortPosR} for vortices with radius that are between $2L$ and $3L$.
\Cref{fig:loop_phi,fig:loop_I} shows how the sizes of the vortex loops depend in superconducting phase difference $\phi$ and magnetic field strength $n$, respectively.
We find that increasing $\phi$ can make the vortices arbitrary large, but does not increase the number of vortices.
Increasing $n$, on the other hand, also increase the number of vortices, but the sizes grow only linearly with $n$.
Note that as the vortex loops hit the surface, they curve so as to hit normally to the surface.
This is consistent with previous results~\cite{Doria2007,romaguero2007}, and can be understood from the circulating currents.
There should be no current component normal to the surface, and the only way for the current circulating the vortices to adhere to this is if the vortices hit the surfaces at a right angle.

\section{Experimental Realization}
\label{sec:expRelWire}
Normal SNS junctions are created by vertically growing first a superconducting material such as niobium, then a normal metal such as copper and finally the same superconducting material.
The layers are grown for instance by a sputter deposition technique such as direct current sputtering~\cite{miller1993} or radio frequency sputtering~\cite{krasnov1994}.
The setup presented here adds an extra complication by requiring an isolated conducting nanowire to penetrate the system.
One possible way to achieve this could be to first grow a vertical insulated nanowire and then grow the superconductor and normal metal around it in a layerwise fashion.

Growing a wire is more complicated than growing a plane because one must localize the growth to happen at the tip at the wire, even though most of the surface area will be on the sides.
Nevertheless, growing vertical nanowires has successfully been done by methods such as the vapor-liquid-solid method~\cite{ng2004,schmidt2006,tomioka2012} and template-directed synthesis~\cite{xia2003}.
The vapor-liquid-solid method works by using droplets of, for instance, gold which are a few angstroms in width to localize the growth~\cite{schmidt2006}, and temple-directed synthesis works by having the wire grow inside a premade template which can later be removed~\cite{xia2003}.
The vapor-liquid-solid method has already been used to produce vertical surround-gate field-effect transistors with a precision exceeding what should be necessary for the system presented here~\cite{schmidt2006}.

\Citet{schmidt2006} made nanowires using the vapor-liquid-solid method which were $\SI{40}{\nano\meter}$ is diameter and $\SI{400}{\nano\meter}$ in length.
This should be on the same length scale as would be necessary for the system considered in this manuscript.
The superconducting energy gap of niobium is $\abs{\Delta} = \SI{30.5e-4}{\electronvolt}$~\cite{kittel1976}, which is equivalent to about $\SI{2.46}{\per\milli\meter}$ in natural units.
The Fermi velocity and scattering time for copper are about $v_F = \SI{3.70e-3}{}$ and $\tau = \SI{10.8}{\micro\meter}$, respectively~\cite{gall2016}.
The diffusion coefficient is defined as
\begin{equation}
  D \defeq \frac{\tau v_F^2}{3},
\end{equation}
so the diffusion coefficient for copper is about $D=\SI{49.2}{\pico\meter}$.
In the numerics we have used
\begin{equation}
  \abs{\Delta} = 4\thouless = \frac{4D}{L^2},
\end{equation}
so
\begin{equation}
  L = \SI{283}{\nano\meter},
\end{equation}
which is on the same scale as what has been made with the vapor-liquid-solid method.
Of course, other metals and superconductors could be used, giving different physical lengths corresponding to the values being used in the numerics here.
Moreover, from the analysis it seems vortex loops would form also for other values of $\abs{\Delta}/\thouless$.
The calculation above is merely to show that the length scales used here is not unreasonable compared to what has already been experimentally achieved.

\section{Conclusion}
We have used quasiclassical Usadel theory to demonstrate that controllable superconducting vortex loops can emerge in a Josephson junction pierced by an insulated current-carrying wire.
The size and number of vortices depend on the phase difference between the superconducting order parameter in the superconductors, $\phi_r - \phi_l$, as well as the strength of the magnetic field.
The radius of the vortices can be made arbitrarily large by tuning of the superconducting phase difference, which means that they can always be manipulated so that they intersect the surface.
This makes them directly observable by scanning tunelling microscopy, which has already been used to detect normal vortices in proximized metals~\cite{Stolyarov2018}.
If this ability of the superconducting phase difference to expand vortex loops to arbitrary sizes is a general feature of SNS-junctions, it could be used to detect vortex loops in systems where controlling the magnetic field strength is not an option, such as in system with a magnetic dipole inclusion.


\begin{acknowledgments}
  We thank M. Amundsen and V. Risinggård for helpful discussions. This work was supported by the Research Council of Norway through grant 240806, and its Centres of Excellence funding scheme grant 262633 ``\emph{QuSpin}''.
\end{acknowledgments}



\newpage
\bibliography{bibliography}

\begin{thebibliography}{39}%
\makeatletter
\providecommand \@ifxundefined [1]{%
 \@ifx{#1\undefined}
}%
\providecommand \@ifnum [1]{%
 \ifnum #1\expandafter \@firstoftwo
 \else \expandafter \@secondoftwo
 \fi
}%
\providecommand \@ifx [1]{%
 \ifx #1\expandafter \@firstoftwo
 \else \expandafter \@secondoftwo
 \fi
}%
\providecommand \natexlab [1]{#1}%
\providecommand \enquote  [1]{``#1''}%
\providecommand \bibnamefont  [1]{#1}%
\providecommand \bibfnamefont [1]{#1}%
\providecommand \citenamefont [1]{#1}%
\providecommand \href@noop [0]{\@secondoftwo}%
\providecommand \href [0]{\begingroup \@sanitize@url \@href}%
\providecommand \@href[1]{\@@startlink{#1}\@@href}%
\providecommand \@@href[1]{\endgroup#1\@@endlink}%
\providecommand \@sanitize@url [0]{\catcode `\\12\catcode `\$12\catcode
  `\&12\catcode `\#12\catcode `\^12\catcode `\_12\catcode `\%12\relax}%
\providecommand \@@startlink[1]{}%
\providecommand \@@endlink[0]{}%
\providecommand \url  [0]{\begingroup\@sanitize@url \@url }%
\providecommand \@url [1]{\endgroup\@href {#1}{\urlprefix }}%
\providecommand \urlprefix  [0]{URL }%
\providecommand \Eprint [0]{\href }%
\providecommand \doibase [0]{http://dx.doi.org/}%
\providecommand \selectlanguage [0]{\@gobble}%
\providecommand \bibinfo  [0]{\@secondoftwo}%
\providecommand \bibfield  [0]{\@secondoftwo}%
\providecommand \translation [1]{[#1]}%
\providecommand \BibitemOpen [0]{}%
\providecommand \bibitemStop [0]{}%
\providecommand \bibitemNoStop [0]{.\EOS\space}%
\providecommand \EOS [0]{\spacefactor3000\relax}%
\providecommand \BibitemShut  [1]{\csname bibitem#1\endcsname}%
\let\auto@bib@innerbib\@empty
\bibitem [{\citenamefont {Fedorov}\ \emph {et~al.}(2014)\citenamefont
  {Fedorov}, \citenamefont {Shcherbakova}, \citenamefont {Wolf}, \citenamefont
  {Beckmann},\ and\ \citenamefont {Ustinov}}]{Fedorov2014}%
  \BibitemOpen
  \bibfield  {author} {\bibinfo {author} {\bibfnamefont {K.~G.}\ \bibnamefont
  {Fedorov}}, \bibinfo {author} {\bibfnamefont {A.~V.}\ \bibnamefont
  {Shcherbakova}}, \bibinfo {author} {\bibfnamefont {M.~J.}\ \bibnamefont
  {Wolf}}, \bibinfo {author} {\bibfnamefont {D.}~\bibnamefont {Beckmann}}, \
  and\ \bibinfo {author} {\bibfnamefont {A.~V.}\ \bibnamefont {Ustinov}},\
  }\href {\doibase 10.1103/PhysRevLett.112.160502} {\bibfield  {journal}
  {\bibinfo  {journal} {Phys. Rev. Lett.}\ }\textbf {\bibinfo {volume} {112}},\
  \bibinfo {pages} {160502} (\bibinfo {year} {2014})}\BibitemShut {NoStop}%
\bibitem [{\citenamefont {Devoret}\ and\ \citenamefont
  {Schoelkopf}(2013)}]{Devoret2013}%
  \BibitemOpen
  \bibfield  {author} {\bibinfo {author} {\bibfnamefont {M.~H.}\ \bibnamefont
  {Devoret}}\ and\ \bibinfo {author} {\bibfnamefont {R.~J.}\ \bibnamefont
  {Schoelkopf}},\ }\href {\doibase 10.1126/science.1231930} {\bibfield
  {journal} {\bibinfo  {journal} {Science}\ }\textbf {\bibinfo {volume}
  {339}},\ \bibinfo {pages} {1169} (\bibinfo {year} {2013})}\BibitemShut
  {NoStop}%
\bibitem [{\citenamefont {Herr}\ \emph {et~al.}(2007)\citenamefont {Herr},
  \citenamefont {Fedorov}, \citenamefont {Shnirman}, \citenamefont {Il'Ichev},\
  and\ \citenamefont {Sch{\"{o}}n}}]{Herr2007}%
  \BibitemOpen
  \bibfield  {author} {\bibinfo {author} {\bibfnamefont {A.}~\bibnamefont
  {Herr}}, \bibinfo {author} {\bibfnamefont {A.}~\bibnamefont {Fedorov}},
  \bibinfo {author} {\bibfnamefont {A.}~\bibnamefont {Shnirman}}, \bibinfo
  {author} {\bibfnamefont {E.}~\bibnamefont {Il'Ichev}}, \ and\ \bibinfo
  {author} {\bibfnamefont {G.}~\bibnamefont {Sch{\"{o}}n}},\ }\href {\doibase
  10.1088/0953-2048/20/11/S29} {\bibfield  {journal} {\bibinfo  {journal}
  {Supercond. Sci. Technol.}\ }\textbf {\bibinfo {volume} {20}},\ \bibinfo
  {pages} {450} (\bibinfo {year} {2007})}\BibitemShut {NoStop}%
\bibitem [{\citenamefont {Kim}\ \emph {et~al.}(2018)\citenamefont {Kim},
  \citenamefont {Myers},\ and\ \citenamefont {Tserkovnyak}}]{Kim2018}%
  \BibitemOpen
  \bibfield  {author} {\bibinfo {author} {\bibfnamefont {S.~K.}\ \bibnamefont
  {Kim}}, \bibinfo {author} {\bibfnamefont {R.}~\bibnamefont {Myers}}, \ and\
  \bibinfo {author} {\bibfnamefont {Y.}~\bibnamefont {Tserkovnyak}},\ }\href
  {\doibase 10.1103/PhysRevLett.121.187203} {\bibfield  {journal} {\bibinfo
  {journal} {Phys. Rev. Lett.}\ }\textbf {\bibinfo {volume} {121}},\ \bibinfo
  {pages} {187203} (\bibinfo {year} {2018})}\BibitemShut {NoStop}%
\bibitem [{\citenamefont {Kwok}\ \emph {et~al.}(2016)\citenamefont {Kwok},
  \citenamefont {Welp}, \citenamefont {Glatz}, \citenamefont {Koshelev},
  \citenamefont {Kihlstrom},\ and\ \citenamefont {Crabtree}}]{Kwok2016}%
  \BibitemOpen
  \bibfield  {author} {\bibinfo {author} {\bibfnamefont {W.~K.}\ \bibnamefont
  {Kwok}}, \bibinfo {author} {\bibfnamefont {U.}~\bibnamefont {Welp}}, \bibinfo
  {author} {\bibfnamefont {A.}~\bibnamefont {Glatz}}, \bibinfo {author}
  {\bibfnamefont {A.~E.}\ \bibnamefont {Koshelev}}, \bibinfo {author}
  {\bibfnamefont {K.~J.}\ \bibnamefont {Kihlstrom}}, \ and\ \bibinfo {author}
  {\bibfnamefont {G.~W.}\ \bibnamefont {Crabtree}},\ }\href {\doibase
  10.1088/0034-4885/79/11/116501} {\bibfield  {journal} {\bibinfo  {journal}
  {Reports Prog. Phys.}\ }\textbf {\bibinfo {volume} {79}},\ \bibinfo {pages}
  {116501} (\bibinfo {year} {2016})}\BibitemShut {NoStop}%
\bibitem [{\citenamefont {{Doria, M. M. and Romaguera, A. R.De C. and
  Milo{\v{s}}evi{\'{c}}, M. V. and Peeters, F. M.}}(2007)}]{Doria2007}%
  \BibitemOpen
  \bibfield  {author} {\bibinfo {author} {\bibnamefont {{Doria, M. M. and
  Romaguera, A. R.De C. and Milo{\v{s}}evi{\'{c}}, M. V. and Peeters, F.
  M.}}},\ }\href {\doibase 10.1209/0295-5075/79/47006} {\bibfield  {journal}
  {\bibinfo  {journal} {Epl}\ }\textbf {\bibinfo {volume} {79}},\ \bibinfo
  {pages} {47006} (\bibinfo {year} {2007})}\BibitemShut {NoStop}%
\bibitem [{\citenamefont {Samokhvalov}(1998)}]{samokhvalov1998}%
  \BibitemOpen
  \bibfield  {author} {\bibinfo {author} {\bibfnamefont {A.~V.}\ \bibnamefont
  {Samokhvalov}},\ }\href {\doibase 10.1016/S0921-4534(98)00416-X} {\bibfield
  {journal} {\bibinfo  {journal} {Physica C}\ }\textbf {\bibinfo {volume}
  {308}},\ \bibinfo {pages} {74 } (\bibinfo {year} {1998})}\BibitemShut
  {NoStop}%
\bibitem [{\citenamefont {Samokhvalov}(1997)}]{samokhvalov1997}%
  \BibitemOpen
  \bibfield  {author} {\bibinfo {author} {\bibfnamefont {A.}~\bibnamefont
  {Samokhvalov}},\ }\href {\doibase 10.1016/S0921-4534(97)01210-0} {\bibfield
  {journal} {\bibinfo  {journal} {Physica C}\ }\textbf {\bibinfo {volume}
  {282-287}},\ \bibinfo {pages} {2163 } (\bibinfo {year} {1997})}\BibitemShut
  {NoStop}%
\bibitem [{\citenamefont {Kozlov}\ and\ \citenamefont
  {Samokhvalov}(1993)}]{kozlov1993}%
  \BibitemOpen
  \bibfield  {author} {\bibinfo {author} {\bibfnamefont {V.}~\bibnamefont
  {Kozlov}}\ and\ \bibinfo {author} {\bibfnamefont {A.}~\bibnamefont
  {Samokhvalov}},\ }\href {\doibase 10.1016/0921-4534(93)90764-H} {\bibfield
  {journal} {\bibinfo  {journal} {Physica C}\ }\textbf {\bibinfo {volume}
  {213}},\ \bibinfo {pages} {103 } (\bibinfo {year} {1993})}\BibitemShut
  {NoStop}%
\bibitem [{\citenamefont {Glatz}\ \emph {et~al.}(2016)\citenamefont {Glatz},
  \citenamefont {Vlasko-Vlasov}, \citenamefont {Kwok},\ and\ \citenamefont
  {Crabtree}}]{Glatz2016}%
  \BibitemOpen
  \bibfield  {author} {\bibinfo {author} {\bibfnamefont {A.}~\bibnamefont
  {Glatz}}, \bibinfo {author} {\bibfnamefont {V.~K.}\ \bibnamefont
  {Vlasko-Vlasov}}, \bibinfo {author} {\bibfnamefont {W.~K.}\ \bibnamefont
  {Kwok}}, \ and\ \bibinfo {author} {\bibfnamefont {G.~W.}\ \bibnamefont
  {Crabtree}},\ }\href {\doibase 10.1103/PhysRevB.94.064505} {\bibfield
  {journal} {\bibinfo  {journal} {Phys. Rev. B}\ }\textbf {\bibinfo {volume}
  {94}},\ \bibinfo {pages} {064505} (\bibinfo {year} {2016})}\BibitemShut
  {NoStop}%
\bibitem [{\citenamefont {Berdiyorov}\ \emph {et~al.}(2018)\citenamefont
  {Berdiyorov}, \citenamefont {Milo{\v{s}}evi{\'{c}}}, \citenamefont
  {Kusmartsev}, \citenamefont {Peeters},\ and\ \citenamefont
  {Savel'ev}}]{Berdiyorov2018}%
  \BibitemOpen
  \bibfield  {author} {\bibinfo {author} {\bibfnamefont {G.~R.}\ \bibnamefont
  {Berdiyorov}}, \bibinfo {author} {\bibfnamefont {M.~V.}\ \bibnamefont
  {Milo{\v{s}}evi{\'{c}}}}, \bibinfo {author} {\bibfnamefont {F.}~\bibnamefont
  {Kusmartsev}}, \bibinfo {author} {\bibfnamefont {F.~M.}\ \bibnamefont
  {Peeters}}, \ and\ \bibinfo {author} {\bibfnamefont {S.}~\bibnamefont
  {Savel'ev}},\ }\href {\doibase 10.1038/s41598-018-21015-7} {\bibfield
  {journal} {\bibinfo  {journal} {Sci. Rep.}\ }\textbf {\bibinfo {volume}
  {8}},\ \bibinfo {pages} {2733} (\bibinfo {year} {2018})}\BibitemShut
  {NoStop}%
\bibitem [{\citenamefont {Sch{\"{o}}nenberger}\ \emph
  {et~al.}(1996)\citenamefont {Sch{\"{o}}nenberger}, \citenamefont {Larkin},
  \citenamefont {Heeb}, \citenamefont {Geshkenbein},\ and\ \citenamefont
  {Blatter}}]{Schonenberger1996}%
  \BibitemOpen
  \bibfield  {author} {\bibinfo {author} {\bibfnamefont {A.}~\bibnamefont
  {Sch{\"{o}}nenberger}}, \bibinfo {author} {\bibfnamefont {A.}~\bibnamefont
  {Larkin}}, \bibinfo {author} {\bibfnamefont {E.}~\bibnamefont {Heeb}},
  \bibinfo {author} {\bibfnamefont {V.}~\bibnamefont {Geshkenbein}}, \ and\
  \bibinfo {author} {\bibfnamefont {G.}~\bibnamefont {Blatter}},\ }\href
  {\doibase 10.1103/PhysRevLett.77.4636} {\bibfield  {journal} {\bibinfo
  {journal} {Phys. Rev. Lett.}\ }\textbf {\bibinfo {volume} {77}},\ \bibinfo
  {pages} {4636} (\bibinfo {year} {1996})}\BibitemShut {NoStop}%
\bibitem [{\citenamefont {Cuevas}\ and\ \citenamefont
  {Bergeret}(2007)}]{Cuevas2007}%
  \BibitemOpen
  \bibfield  {author} {\bibinfo {author} {\bibfnamefont {J.~C.}\ \bibnamefont
  {Cuevas}}\ and\ \bibinfo {author} {\bibfnamefont {F.~S.}\ \bibnamefont
  {Bergeret}},\ }\href {\doibase 10.1103/PhysRevLett.99.217002} {\bibfield
  {journal} {\bibinfo  {journal} {Phys. Rev. Lett.}\ }\textbf {\bibinfo
  {volume} {99}},\ \bibinfo {pages} {217002} (\bibinfo {year}
  {2007})}\BibitemShut {NoStop}%
\bibitem [{\citenamefont {Stolyarov}\ \emph {et~al.}(2018)\citenamefont
  {Stolyarov}, \citenamefont {Cren}, \citenamefont {Brun}, \citenamefont
  {Golovchanskiy}, \citenamefont {Skryabina}, \citenamefont {Kasatonov},
  \citenamefont {Khapaev}, \citenamefont {Kupriyanov}, \citenamefont
  {Golubov},\ and\ \citenamefont {Roditchev}}]{Stolyarov2018}%
  \BibitemOpen
  \bibfield  {author} {\bibinfo {author} {\bibfnamefont {V.~S.}\ \bibnamefont
  {Stolyarov}}, \bibinfo {author} {\bibfnamefont {T.}~\bibnamefont {Cren}},
  \bibinfo {author} {\bibfnamefont {C.}~\bibnamefont {Brun}}, \bibinfo {author}
  {\bibfnamefont {I.~A.}\ \bibnamefont {Golovchanskiy}}, \bibinfo {author}
  {\bibfnamefont {O.~V.}\ \bibnamefont {Skryabina}}, \bibinfo {author}
  {\bibfnamefont {D.~I.}\ \bibnamefont {Kasatonov}}, \bibinfo {author}
  {\bibfnamefont {M.~M.}\ \bibnamefont {Khapaev}}, \bibinfo {author}
  {\bibfnamefont {M.~Y.}\ \bibnamefont {Kupriyanov}}, \bibinfo {author}
  {\bibfnamefont {A.~A.}\ \bibnamefont {Golubov}}, \ and\ \bibinfo {author}
  {\bibfnamefont {D.}~\bibnamefont {Roditchev}},\ }\href {\doibase
  10.1038/s41467-018-04582-1} {\bibfield  {journal} {\bibinfo  {journal} {Nat.
  Commun.}\ }\textbf {\bibinfo {volume} {9}},\ \bibinfo {pages} {2277}
  (\bibinfo {year} {2018})}\BibitemShut {NoStop}%
\bibitem [{\citenamefont {Usadel}(1970)}]{Usadel1970}%
  \BibitemOpen
  \bibfield  {author} {\bibinfo {author} {\bibfnamefont {K.~D.}\ \bibnamefont
  {Usadel}},\ }\href {\doibase 10.1103/PhysRevLett.25.507} {\bibfield
  {journal} {\bibinfo  {journal} {Phys. Rev. Lett.}\ }\textbf {\bibinfo
  {volume} {25}},\ \bibinfo {pages} {507} (\bibinfo {year} {1970})}\BibitemShut
  {NoStop}%
\bibitem [{\citenamefont {Zaikin}\ \emph {et~al.}(2002)\citenamefont {Zaikin},
  \citenamefont {Bruder}, \citenamefont {Belzig}, \citenamefont {Sch{\"{o}}n},\
  and\ \citenamefont {Wilhelm}}]{Belzig1999}%
  \BibitemOpen
  \bibfield  {author} {\bibinfo {author} {\bibfnamefont {A.~D.}\ \bibnamefont
  {Zaikin}}, \bibinfo {author} {\bibfnamefont {C.}~\bibnamefont {Bruder}},
  \bibinfo {author} {\bibfnamefont {W.}~\bibnamefont {Belzig}}, \bibinfo
  {author} {\bibfnamefont {G.}~\bibnamefont {Sch{\"{o}}n}}, \ and\ \bibinfo
  {author} {\bibfnamefont {F.~K.}\ \bibnamefont {Wilhelm}},\ }\href {\doibase
  10.1006/spmi.1999.0710} {\bibfield  {journal} {\bibinfo  {journal}
  {Superlattices Microstruct.}\ }\textbf {\bibinfo {volume} {25}},\ \bibinfo
  {pages} {1251} (\bibinfo {year} {2002})}\BibitemShut {NoStop}%
\bibitem [{\citenamefont {Chandrasekhar}(2004)}]{Chandrasekhar2004}%
  \BibitemOpen
  \bibfield  {author} {\bibinfo {author} {\bibfnamefont {V.}~\bibnamefont
  {Chandrasekhar}},\ }in\ \href {\doibase 10.1007/978-3-642-18914-2_3} {\emph
  {\bibinfo {booktitle} {Phys. Supercond.}}}\ (\bibinfo  {publisher} {Springer
  Berlin Heidelberg},\ \bibinfo {address} {Berlin, Heidelberg},\ \bibinfo
  {year} {2004})\ pp.\ \bibinfo {pages} {55--110}\BibitemShut {NoStop}%
\bibitem [{\citenamefont {Rammer}(2004)}]{Rammer2004}%
  \BibitemOpen
  \bibfield  {author} {\bibinfo {author} {\bibfnamefont {J.}~\bibnamefont
  {Rammer}},\ }\href
  {https://www.crcpress.com/Quantum-Transport-Theory/Rammer/p/book/9780813342849}
  {\emph {\bibinfo {title} {{Quantum transport theory}}}}\ (\bibinfo
  {publisher} {Westview},\ \bibinfo {year} {2004})\ p.\ \bibinfo {pages}
  {521}\BibitemShut {NoStop}%
\bibitem [{\citenamefont {Barone}\ and\ \citenamefont
  {Paternò}(1982)}]{Barone1982}%
  \BibitemOpen
  \bibfield  {author} {\bibinfo {author} {\bibfnamefont {A.}~\bibnamefont
  {Barone}}\ and\ \bibinfo {author} {\bibfnamefont {G.}~\bibnamefont
  {Paternò}},\ }\href {http://cds.cern.ch/record/101094} {\emph {\bibinfo
  {title} {{Physics and applications of the Josephson effect}}}}\ (\bibinfo
  {publisher} {Wiley},\ \bibinfo {address} {New York, NY},\ \bibinfo {year}
  {1982})\BibitemShut {NoStop}%
\bibitem [{\citenamefont {Alidoust}\ and\ \citenamefont
  {Halterman}(2015)}]{Alidoust2015}%
  \BibitemOpen
  \bibfield  {author} {\bibinfo {author} {\bibfnamefont {M.}~\bibnamefont
  {Alidoust}}\ and\ \bibinfo {author} {\bibfnamefont {K.}~\bibnamefont
  {Halterman}},\ }\href {\doibase 10.1063/1.4908287} {\bibfield  {journal}
  {\bibinfo  {journal} {J. Appl. Phys.}\ }\textbf {\bibinfo {volume} {117}},\
  \bibinfo {pages} {123906} (\bibinfo {year} {2015})}\BibitemShut {NoStop}%
\bibitem [{\citenamefont {Bergeret}\ and\ \citenamefont
  {Cuevas}(2008)}]{Bergeret2008a}%
  \BibitemOpen
  \bibfield  {author} {\bibinfo {author} {\bibfnamefont {F.~S.}\ \bibnamefont
  {Bergeret}}\ and\ \bibinfo {author} {\bibfnamefont {J.~C.}\ \bibnamefont
  {Cuevas}},\ }\href {\doibase 10.1007/s10909-008-9826-2} {\bibfield  {journal}
  {\bibinfo  {journal} {J. Low Temp. Phys.}\ }\textbf {\bibinfo {volume}
  {153}},\ \bibinfo {pages} {304} (\bibinfo {year} {2008})}\BibitemShut
  {NoStop}%
\bibitem [{\citenamefont {Chiodi}\ \emph {et~al.}(2012)\citenamefont {Chiodi},
  \citenamefont {Ferrier}, \citenamefont {Gu\'eron}, \citenamefont {Cuevas},
  \citenamefont {Montambaux}, \citenamefont {Fortuna}, \citenamefont
  {Kasumov},\ and\ \citenamefont {Bouchiat}}]{chiodi2012}%
  \BibitemOpen
  \bibfield  {author} {\bibinfo {author} {\bibfnamefont {F.}~\bibnamefont
  {Chiodi}}, \bibinfo {author} {\bibfnamefont {M.}~\bibnamefont {Ferrier}},
  \bibinfo {author} {\bibfnamefont {S.}~\bibnamefont {Gu\'eron}}, \bibinfo
  {author} {\bibfnamefont {J.~C.}\ \bibnamefont {Cuevas}}, \bibinfo {author}
  {\bibfnamefont {G.}~\bibnamefont {Montambaux}}, \bibinfo {author}
  {\bibfnamefont {F.}~\bibnamefont {Fortuna}}, \bibinfo {author} {\bibfnamefont
  {A.}~\bibnamefont {Kasumov}}, \ and\ \bibinfo {author} {\bibfnamefont
  {H.}~\bibnamefont {Bouchiat}},\ }\href {\doibase 10.1103/PhysRevB.86.064510}
  {\bibfield  {journal} {\bibinfo  {journal} {Phys. Rev. B}\ }\textbf {\bibinfo
  {volume} {86}},\ \bibinfo {pages} {064510} (\bibinfo {year}
  {2012})}\BibitemShut {NoStop}%
\bibitem [{\citenamefont {Schopohl}(1998)}]{schopohl_arxiv_98}%
  \BibitemOpen
  \bibfield  {author} {\bibinfo {author} {\bibfnamefont {N.}~\bibnamefont
  {Schopohl}},\ }\href {https://arxiv.org/pdf/cond-mat/9804064.pdf} {\bibfield
  {journal} {\bibinfo  {journal} {arXiv:cond-mat/9804064}\ } (\bibinfo {year}
  {1998})}\BibitemShut {NoStop}%
\bibitem [{\citenamefont {Amundsen}\ and\ \citenamefont
  {Linder}(2016)}]{Amundsen2016}%
  \BibitemOpen
  \bibfield  {author} {\bibinfo {author} {\bibfnamefont {M.}~\bibnamefont
  {Amundsen}}\ and\ \bibinfo {author} {\bibfnamefont {J.}~\bibnamefont
  {Linder}},\ }\href {\doibase 10.1038/srep22765} {\bibfield  {journal}
  {\bibinfo  {journal} {Sci. Rep.}\ }\textbf {\bibinfo {volume} {6}},\ \bibinfo
  {pages} {22765} (\bibinfo {year} {2016})}\BibitemShut {NoStop}%
\bibitem [{\citenamefont {Bezanson}\ \emph {et~al.}(2017)\citenamefont
  {Bezanson}, \citenamefont {Edelman}, \citenamefont {Karpinski},\ and\
  \citenamefont {Shah}}]{Bezanson2017}%
  \BibitemOpen
  \bibfield  {author} {\bibinfo {author} {\bibfnamefont {J.}~\bibnamefont
  {Bezanson}}, \bibinfo {author} {\bibfnamefont {A.}~\bibnamefont {Edelman}},
  \bibinfo {author} {\bibfnamefont {S.}~\bibnamefont {Karpinski}}, \ and\
  \bibinfo {author} {\bibfnamefont {V.~B.}\ \bibnamefont {Shah}},\ }\href
  {\doibase 10.1137/141000671} {\bibfield  {journal} {\bibinfo  {journal} {SIAM
  Rev.}\ }\textbf {\bibinfo {volume} {59}},\ \bibinfo {pages} {65} (\bibinfo
  {year} {2017})}\BibitemShut {NoStop}%
\bibitem [{\citenamefont {Carlsson}(2019)}]{Carlsson2019}%
  \BibitemOpen
  \bibfield  {author} {\bibinfo {author} {\bibfnamefont {K.}~\bibnamefont
  {Carlsson}},\ }\href@noop {} {\enquote {\bibinfo {title} {Juafem.jl},}\
  }\bibinfo {howpublished} {\url{https://github.com/KristofferC/JuAFEM.jl}}
  (\bibinfo {year} {2019})\BibitemShut {NoStop}%
\bibitem [{\citenamefont {Sauer}(2013)}]{sauer2013}%
  \BibitemOpen
  \bibfield  {author} {\bibinfo {author} {\bibfnamefont {T.}~\bibnamefont
  {Sauer}},\ }\href@noop {} {\emph {\bibinfo {title} {Numerical Analysis:
  Pearson New International Edition}}}\ (\bibinfo  {publisher} {Pearson
  Education Limited},\ \bibinfo {year} {2013})\BibitemShut {NoStop}%
\bibitem [{\citenamefont {Revels}\ \emph {et~al.}(2016)\citenamefont {Revels},
  \citenamefont {Lubin},\ and\ \citenamefont
  {Papamarkou}}]{RevelsLubinPapamarkou2016}%
  \BibitemOpen
  \bibfield  {author} {\bibinfo {author} {\bibfnamefont {J.}~\bibnamefont
  {Revels}}, \bibinfo {author} {\bibfnamefont {M.}~\bibnamefont {Lubin}}, \
  and\ \bibinfo {author} {\bibfnamefont {T.}~\bibnamefont {Papamarkou}},\
  }\href {https://arxiv.org/abs/1607.07892} {\bibfield  {journal} {\bibinfo
  {journal} {arXiv:1607.07892 [cs.MS]}\ } (\bibinfo {year} {2016})}\BibitemShut
  {NoStop}%
\bibitem [{\citenamefont {Dynes}\ \emph {et~al.}(1984)\citenamefont {Dynes},
  \citenamefont {Garno}, \citenamefont {Hertel},\ and\ \citenamefont
  {Orlando}}]{Dynes1984}%
  \BibitemOpen
  \bibfield  {author} {\bibinfo {author} {\bibfnamefont {R.~C.}\ \bibnamefont
  {Dynes}}, \bibinfo {author} {\bibfnamefont {J.~P.}\ \bibnamefont {Garno}},
  \bibinfo {author} {\bibfnamefont {G.~B.}\ \bibnamefont {Hertel}}, \ and\
  \bibinfo {author} {\bibfnamefont {T.~P.}\ \bibnamefont {Orlando}},\ }\href
  {\doibase 10.1103/PhysRevLett.53.2437} {\bibfield  {journal} {\bibinfo
  {journal} {Phys. Rev. Lett.}\ }\textbf {\bibinfo {volume} {53}},\ \bibinfo
  {pages} {2437} (\bibinfo {year} {1984})}\BibitemShut {NoStop}%
\bibitem [{\citenamefont {Sonin}(1997)}]{sonin1997}%
  \BibitemOpen
  \bibfield  {author} {\bibinfo {author} {\bibfnamefont {E.~B.}\ \bibnamefont
  {Sonin}},\ }\href {\doibase 10.1103/PhysRevB.55.485} {\bibfield  {journal}
  {\bibinfo  {journal} {Phys. Rev. B}\ }\textbf {\bibinfo {volume} {55}},\
  \bibinfo {pages} {485} (\bibinfo {year} {1997})}\BibitemShut {NoStop}%
\bibitem [{\citenamefont {de~C.~Romaguera}\ \emph {et~al.}(2007)\citenamefont
  {de~C.~Romaguera}, \citenamefont {Doria},\ and\ \citenamefont
  {Peeters}}]{romaguero2007}%
  \BibitemOpen
  \bibfield  {author} {\bibinfo {author} {\bibfnamefont {A.~R.}\ \bibnamefont
  {de~C.~Romaguera}}, \bibinfo {author} {\bibfnamefont {M.~M.}\ \bibnamefont
  {Doria}}, \ and\ \bibinfo {author} {\bibfnamefont {F.~M.}\ \bibnamefont
  {Peeters}},\ }\href {\doibase 10.1103/PhysRevB.75.184525} {\bibfield
  {journal} {\bibinfo  {journal} {Phys. Rev. B}\ }\textbf {\bibinfo {volume}
  {75}},\ \bibinfo {pages} {184525} (\bibinfo {year} {2007})}\BibitemShut
  {NoStop}%
\bibitem [{\citenamefont {Miller}\ \emph {et~al.}(1993)\citenamefont {Miller},
  \citenamefont {Mallison}, \citenamefont {Kleinsasser}, \citenamefont
  {Delin},\ and\ \citenamefont {Macedo}}]{miller1993}%
  \BibitemOpen
  \bibfield  {author} {\bibinfo {author} {\bibfnamefont {R.~E.}\ \bibnamefont
  {Miller}}, \bibinfo {author} {\bibfnamefont {W.~H.}\ \bibnamefont
  {Mallison}}, \bibinfo {author} {\bibfnamefont {A.~W.}\ \bibnamefont
  {Kleinsasser}}, \bibinfo {author} {\bibfnamefont {K.~A.}\ \bibnamefont
  {Delin}}, \ and\ \bibinfo {author} {\bibfnamefont {E.~M.}\ \bibnamefont
  {Macedo}},\ }\href {\doibase 10.1063/1.109697} {\bibfield  {journal}
  {\bibinfo  {journal} {Appl. Phys. Lett.}\ }\textbf {\bibinfo {volume} {63}},\
  \bibinfo {pages} {1423} (\bibinfo {year} {1993})}\BibitemShut {NoStop}%
\bibitem [{\citenamefont {Krasnov}\ \emph {et~al.}(1994)\citenamefont
  {Krasnov}, \citenamefont {Pedersen}, \citenamefont {Oboznov},\ and\
  \citenamefont {Ryazanov}}]{krasnov1994}%
  \BibitemOpen
  \bibfield  {author} {\bibinfo {author} {\bibfnamefont {V.~M.}\ \bibnamefont
  {Krasnov}}, \bibinfo {author} {\bibfnamefont {N.~F.}\ \bibnamefont
  {Pedersen}}, \bibinfo {author} {\bibfnamefont {V.~A.}\ \bibnamefont
  {Oboznov}}, \ and\ \bibinfo {author} {\bibfnamefont {V.~V.}\ \bibnamefont
  {Ryazanov}},\ }\href {\doibase 10.1103/PhysRevB.49.12969} {\bibfield
  {journal} {\bibinfo  {journal} {Phys. Rev. B}\ }\textbf {\bibinfo {volume}
  {49}},\ \bibinfo {pages} {12969} (\bibinfo {year} {1994})}\BibitemShut
  {NoStop}%
\bibitem [{\citenamefont {Ng}\ \emph {et~al.}(2004)\citenamefont {Ng},
  \citenamefont {Han}, \citenamefont {Yamada}, \citenamefont {Nguyen},
  \citenamefont {Chen},\ and\ \citenamefont {Meyyappan}}]{ng2004}%
  \BibitemOpen
  \bibfield  {author} {\bibinfo {author} {\bibfnamefont {H.~T.}\ \bibnamefont
  {Ng}}, \bibinfo {author} {\bibfnamefont {J.}~\bibnamefont {Han}}, \bibinfo
  {author} {\bibfnamefont {T.}~\bibnamefont {Yamada}}, \bibinfo {author}
  {\bibfnamefont {P.}~\bibnamefont {Nguyen}}, \bibinfo {author} {\bibfnamefont
  {Y.~P.}\ \bibnamefont {Chen}}, \ and\ \bibinfo {author} {\bibfnamefont
  {M.}~\bibnamefont {Meyyappan}},\ }\href {\doibase 10.1021/nl049461z}
  {\bibfield  {journal} {\bibinfo  {journal} {Nano Lett.}\ }\textbf {\bibinfo
  {volume} {4}},\ \bibinfo {pages} {1247} (\bibinfo {year} {2004})}\BibitemShut
  {NoStop}%
\bibitem [{\citenamefont {Schmidt}\ \emph {et~al.}(2006)\citenamefont
  {Schmidt}, \citenamefont {Riel}, \citenamefont {Senz}, \citenamefont {Karg},
  \citenamefont {Riess},\ and\ \citenamefont {Gösele}}]{schmidt2006}%
  \BibitemOpen
  \bibfield  {author} {\bibinfo {author} {\bibfnamefont {V.}~\bibnamefont
  {Schmidt}}, \bibinfo {author} {\bibfnamefont {H.}~\bibnamefont {Riel}},
  \bibinfo {author} {\bibfnamefont {S.}~\bibnamefont {Senz}}, \bibinfo {author}
  {\bibfnamefont {S.}~\bibnamefont {Karg}}, \bibinfo {author} {\bibfnamefont
  {W.}~\bibnamefont {Riess}}, \ and\ \bibinfo {author} {\bibfnamefont
  {U.}~\bibnamefont {Gösele}},\ }\href {\doibase 10.1002/smll.200500181}
  {\bibfield  {journal} {\bibinfo  {journal} {Small}\ }\textbf {\bibinfo
  {volume} {2}},\ \bibinfo {pages} {85} (\bibinfo {year} {2006})}\BibitemShut
  {NoStop}%
\bibitem [{\citenamefont {Tomioka}\ \emph {et~al.}(2012)\citenamefont
  {Tomioka}, \citenamefont {Yoshimura},\ and\ \citenamefont
  {Fukui}}]{tomioka2012}%
  \BibitemOpen
  \bibfield  {author} {\bibinfo {author} {\bibfnamefont {K.}~\bibnamefont
  {Tomioka}}, \bibinfo {author} {\bibfnamefont {M.}~\bibnamefont {Yoshimura}},
  \ and\ \bibinfo {author} {\bibfnamefont {T.}~\bibnamefont {Fukui}},\ }\href
  {\doibase 10.1038/nature11293} {\bibfield  {journal} {\bibinfo  {journal}
  {Nature}\ }\textbf {\bibinfo {volume} {488}},\ \bibinfo {pages} {189}
  (\bibinfo {year} {2012})}\BibitemShut {NoStop}%
\bibitem [{\citenamefont {Xia}\ \emph {et~al.}(2003)\citenamefont {Xia},
  \citenamefont {Yang}, \citenamefont {Sun}, \citenamefont {Wu}, \citenamefont
  {Mayers}, \citenamefont {Gates}, \citenamefont {Yin}, \citenamefont {Kim},\
  and\ \citenamefont {Yan}}]{xia2003}%
  \BibitemOpen
  \bibfield  {author} {\bibinfo {author} {\bibfnamefont {Y.}~\bibnamefont
  {Xia}}, \bibinfo {author} {\bibfnamefont {P.}~\bibnamefont {Yang}}, \bibinfo
  {author} {\bibfnamefont {Y.}~\bibnamefont {Sun}}, \bibinfo {author}
  {\bibfnamefont {Y.}~\bibnamefont {Wu}}, \bibinfo {author} {\bibfnamefont
  {B.}~\bibnamefont {Mayers}}, \bibinfo {author} {\bibfnamefont
  {B.}~\bibnamefont {Gates}}, \bibinfo {author} {\bibfnamefont
  {Y.}~\bibnamefont {Yin}}, \bibinfo {author} {\bibfnamefont {F.}~\bibnamefont
  {Kim}}, \ and\ \bibinfo {author} {\bibfnamefont {H.}~\bibnamefont {Yan}},\
  }\href {\doibase 10.1002/adma.200390087} {\bibfield  {journal} {\bibinfo
  {journal} {Advanced Materials}\ }\textbf {\bibinfo {volume} {15}},\ \bibinfo
  {pages} {353} (\bibinfo {year} {2003})}\BibitemShut {NoStop}%
\bibitem [{\citenamefont {Kittel}\ \emph {et~al.}(1976)\citenamefont {Kittel}
  \emph {et~al.}}]{kittel1976}%
  \BibitemOpen
  \bibfield  {author} {\bibinfo {author} {\bibfnamefont {C.}~\bibnamefont
  {Kittel}} \emph {et~al.},\ }\href@noop {} {\emph {\bibinfo {title}
  {Introduction to solid state physics}}},\ Vol.~\bibinfo {volume} {8}\
  (\bibinfo  {publisher} {Wiley New York},\ \bibinfo {year} {1976})\BibitemShut
  {NoStop}%
\bibitem [{\citenamefont {Gall}(2016)}]{gall2016}%
  \BibitemOpen
  \bibfield  {author} {\bibinfo {author} {\bibfnamefont {D.}~\bibnamefont
  {Gall}},\ }\href {\doibase 10.1063/1.4942216} {\bibfield  {journal} {\bibinfo
   {journal} {J. Appl. Phys.}\ }\textbf {\bibinfo {volume} {119}},\ \bibinfo
  {pages} {085101} (\bibinfo {year} {2016})}\BibitemShut {NoStop}%
\end{thebibliography}%

\end{document}